\documentclass[aps,prl,twocolumn,superscriptaddress,showpacs,nofootinbib,longbibliography]{revtex4-1}
\usepackage{graphicx}
\usepackage{dcolumn}
\usepackage{bm}
\usepackage{hyperref}
\usepackage{upgreek}
\usepackage{color}
\usepackage{soul}
\usepackage{epstopdf}
\usepackage{units}
\usepackage{longtable}
\usepackage{floatrow}
\usepackage{latexsym}
\usepackage{gensymb}
\usepackage{floatrow}

\begin{document}

\title{Any Axion Insulator Must be a Bulk Three-Dimensional Topological Insulator}

\author{K. M. Fijalkowski}
\affiliation{Faculty for Physics and Astronomy (EP3), Universit\"at W\"urzburg, Am Hubland, D-97074, W\"urzburg, Germany}
\affiliation{Institute for Topological Insulators, Am Hubland, D-97074, W\"urzburg, Germany}
\author{N. Liu}
\affiliation{Faculty for Physics and Astronomy (EP3), Universit\"at W\"urzburg, Am Hubland, D-97074, W\"urzburg, Germany}
\affiliation{Institute for Topological Insulators, Am Hubland, D-97074, W\"urzburg, Germany}
\author{M. Hartl}
\affiliation{Faculty for Physics and Astronomy (EP3), Universit\"at W\"urzburg, Am Hubland, D-97074, W\"urzburg, Germany}
\affiliation{Institute for Topological Insulators, Am Hubland, D-97074, W\"urzburg, Germany}
\author{M. Winnerlein}
\affiliation{Faculty for Physics and Astronomy (EP3), Universit\"at W\"urzburg, Am Hubland, D-97074, W\"urzburg, Germany}
\affiliation{Institute for Topological Insulators, Am Hubland, D-97074, W\"urzburg, Germany}
\author{P. Mandal}
\affiliation{Faculty for Physics and Astronomy (EP3), Universit\"at W\"urzburg, Am Hubland, D-97074, W\"urzburg, Germany}
\affiliation{Institute for Topological Insulators, Am Hubland, D-97074, W\"urzburg, Germany}
\author{A. Coschizza}
\affiliation{Faculty for Physics and Astronomy (EP3), Universit\"at W\"urzburg, Am Hubland, D-97074, W\"urzburg, Germany}
\affiliation{Institute for Topological Insulators, Am Hubland, D-97074, W\"urzburg, Germany}
\author{A. Fothergill}
\affiliation{Faculty for Physics and Astronomy (EP3), Universit\"at W\"urzburg, Am Hubland, D-97074, W\"urzburg, Germany}
\affiliation{Institute for Topological Insulators, Am Hubland, D-97074, W\"urzburg, Germany}
\author{S. Grauer}
\affiliation{Faculty for Physics and Astronomy (EP3), Universit\"at W\"urzburg, Am Hubland, D-97074, W\"urzburg, Germany}
\affiliation{Institute for Topological Insulators, Am Hubland, D-97074, W\"urzburg, Germany}
\author{S. Schreyeck}
\affiliation{Faculty for Physics and Astronomy (EP3), Universit\"at W\"urzburg, Am Hubland, D-97074, W\"urzburg, Germany}
\affiliation{Institute for Topological Insulators, Am Hubland, D-97074, W\"urzburg, Germany}
\author{K. Brunner}
\affiliation{Faculty for Physics and Astronomy (EP3), Universit\"at W\"urzburg, Am Hubland, D-97074, W\"urzburg, Germany}
\affiliation{Institute for Topological Insulators, Am Hubland, D-97074, W\"urzburg, Germany}
\author{M. Greiter}
\affiliation{Faculty for Physics and Astronomy (TP1), Universit\"at W\"urzburg, Am Hubland, D-97074, W\"urzburg, Germany}
\author{R. Thomale}
\affiliation{Faculty for Physics and Astronomy (TP1), Universit\"at W\"urzburg, Am Hubland, D-97074, W\"urzburg, Germany}
\author{C. Gould}
\affiliation{Faculty for Physics and Astronomy (EP3), Universit\"at W\"urzburg, Am Hubland, D-97074, W\"urzburg, Germany}
\affiliation{Institute for Topological Insulators, Am Hubland, D-97074, W\"urzburg, Germany}
\author{L. W. Molenkamp}
\affiliation{Faculty for Physics and Astronomy (EP3), Universit\"at W\"urzburg, Am Hubland, D-97074, W\"urzburg, Germany}
\affiliation{Institute for Topological Insulators, Am Hubland, D-97074, W\"urzburg, Germany}

\date{\today}

\begin{abstract}
{In recent attempts to observe axion electrodynamics, much effort has focused on trilayer heterostructures of magnetic topological insulators, and in particular on the examination of a so-called zero Hall plateau, which has misguidedly been overstated as direct evidence of an axion insulator state. We investigate the general notion of axion insulators, which by definition must contain a non-trivial volume to host the axion term. We conduct a detailed magneto-transport analysis of Chern insulators comprised of a single magnetic topological insulator layer of varying thickness as well as trilayer structures, for samples optimized to yield a perfectly quantized anomalous Hall effect. Our analysis gives evidence for a topological magneto-electric effect quantized in units of $e^2/2h$, allowing us to identify signatures of axion electrodynamics. Our observations may provide direct experimental access to electrodynamic properties of the universe beyond the traditional Maxwell equations, and challenge the hitherto proclaimed exclusive link between the observation of a zero Hall plateau and an axion insulator.}
\end{abstract}
\maketitle

Topological insulators (TIs) are now established as a new paradigm for quantum states of matter. From the beginning, the field has faced a challenge in adjusting its nomenclature to the increasing diversification of experimental evidence. The prediction and subsequent discovery of the quantum spin Hall effect~\cite{Kanemele1,Kanemele2,Bernevig2006,Konig2007} established the notion of a topological insulator defined by an insulating two dimensional (2D) bulk, a non-trivial Z$_2$ topological bulk invariant, and a corresponding 1D edge state along the bulk boundary. In theoretical considerations, the preferred perspective avoids edges altogether by assuming periodic boundary conditions in calculating the topological bulk invariant. It is however the edge states that reveal the quantum spin Hall effect through transport experiments~\cite{Konig2007}. 

The meaning of the expression topological insulator~\cite{PhysRevB.75.121306} was extended when parity band inversion induced by spin orbit coupling was theoretically conceptualized for arbitrary spatial dimensions ~\cite{PhysRevLett.98.106803,PhysRevB.76.045302} and then subsequently realized in three-dimensional crystals~\cite{Bruene2011}. As in the 2D case, a (strong) 3D topological insulator is characterized by an insulating 3D bulk, a non-trivial Z$_2$ 3D bulk invariant, and usually a single (or in general an odd number of) Dirac cones as its surface states. Unlike the 2D case, identifying the transport signature of 3D topological insulators has remained challenging, with indications of surface state transport found in the quantum Hall signals~\cite{Bruene2011,Xu2014}, whereas the spectroscopic identification of the surface Dirac cones turned out to be comparably straightforward~\cite{hasannature}. More than just a non-trivial bulk invariant, however, a 3D topological insulator also alters its electromagnetic response in the vacuum formed by the electronic insulator~\cite{PhysRevB.78.195424,PhysRevLett.102.146805}. The action term takes the form
\begin{equation}
S=\frac{\theta e^2}{4\pi^2\hbar}\int d^3x dt \; {\bf E} \cdot {\bf B}
 \end{equation}
where $\theta$ denotes the axion field~\cite{PhysRevLett.58.1799} constrained to take values of $0$ or $\pi$ for time reversal symmetric systems, and ${\bf E}, {\bf B}$ the electric and magnetic field, respectively. In the bulk of a 3D TI, $\theta=\pi$, with a $\theta$-field gradient setting in near domain walls or surface termination where the value of $\theta$ changes, such as for example when the value of $\theta$ changes from $\pi$ to $0$ at the interface between the TI bulk and the time reversal symmetric topologically trivial vacuum.

At this point, one might be tempted to speculate that the wording axion insulator could be used synonymously to a 3D strong topological insulator. Instead, a variety of largely confusing definitions have been proliferating in the literature. Most of the time, the definition axion insulator appears to have an operational background, emphasizing that one wants to prepare a system in which the axion electrodynamics can be measured. For that purpose, the signal stemming from eq. (1) should not be obscured by responses from the surface states. As a consequence, an axion insulator often is just defined as a topological insulator with an axion electromagnetic response and all surface states gapped, preferably through the breaking of time-reversal symmetry solely at the surface, and not the bulk. The multiple shortcomings of this wording become even more apparent when analyzing systems which have been labeled trilayer axion insulators~\cite{PhysRevB.92.081107,PhysRevB.92.085113}: A thin layer of a non-magnetic TI is sandwiched by two magnetic TI layers with a magnetic field pointing outwards along the surface normal and assumed to gap out the electronic states at the surfaces. The absence of any surface state transport can be verified by measuring a large $\rho_{xx}$, whereas $\sigma_{xy}$ is expected to cancel over the two surfaces because of global time reversal symmetry. In this logic, the magnetic field orientation in the upper and lower magnetic TI layers has to be opposite. Following this argument, the observation of a zero Hall plateau has been mistakenly interpreted as a sufficient condition for a system in which the axion term $\theta$ can be addressed experimentally~\cite{PhysRevLett.120.056801,tokura1,Mogi20172,Liu2020}.

As we discuss more thoroughly later, such TI hetero-structure samples lack 3D bulk properties. Despite this, it has been claimed that they can still be a source for phenomena stemming from eq. (1). The narrative is that the Lagrangian density of (1) can be rewritten as a total derivative and thus reduced to a surface action $S=\theta e^2/8\pi^2\hbar \int d^2x dt \epsilon^{\mu\nu\lambda}A_\mu \partial_\nu A_\lambda$.  This manifests as a half integer Chern-Simons term, yielding the topological magneto-electric effect~\cite{PhysRevB.78.195424} quantized in units of $e^2/2h$. While it is an open task and of course desirable to resolve this quantization effect in experiment, one could argue that the natural presence of this term at the surface of {\it any} 3D TI is already enforced by gauge symmetry for an odd number of surface Dirac cones~\cite{PhysRevLett.52.18}. More critically, however, the reduction of (1) to a surface Chern-Simons term necessitates a valid continuum distinction between bulk and surface, which as we will demonstrate is not the case for trilayers. A zero Hall plateau together with the insulating behavior inferred from a large longitudinal resistance, which have often been attributed to axion insulator states, are not unique to magnetic hetero-structures. Indeed, from the point of view of electrodynamics, the TI trilayers are experimentally indistinguishable from ordinary 2D systems without any axion physics.

In this manuscript we show, by pointing to several ambiguities surrounding the zero Hall plateau axion insulator interdependence, that currently trending ideas for an "axion insulator" require revision. We thus propose a most basic and minimal definition for an "axion insulator", namely: "An axion insulator is a bulk three-dimensional insulator supporting a quantized axion term within its volume". This definition not only acknowledges the initial theoretical narrative towards existence of axion physics in three-dimensional topological insulators \cite{PhysRevB.78.195424,Qi2011}, but also removes the incorrectly proclaimed one to one correspondence between axion electrodynamics and a zero Hall plateau.

\onecolumngrid
\noindent\rule[0.5ex]{\linewidth}{1pt}

\begin{figure}[h]
\includegraphics[width=\columnwidth]{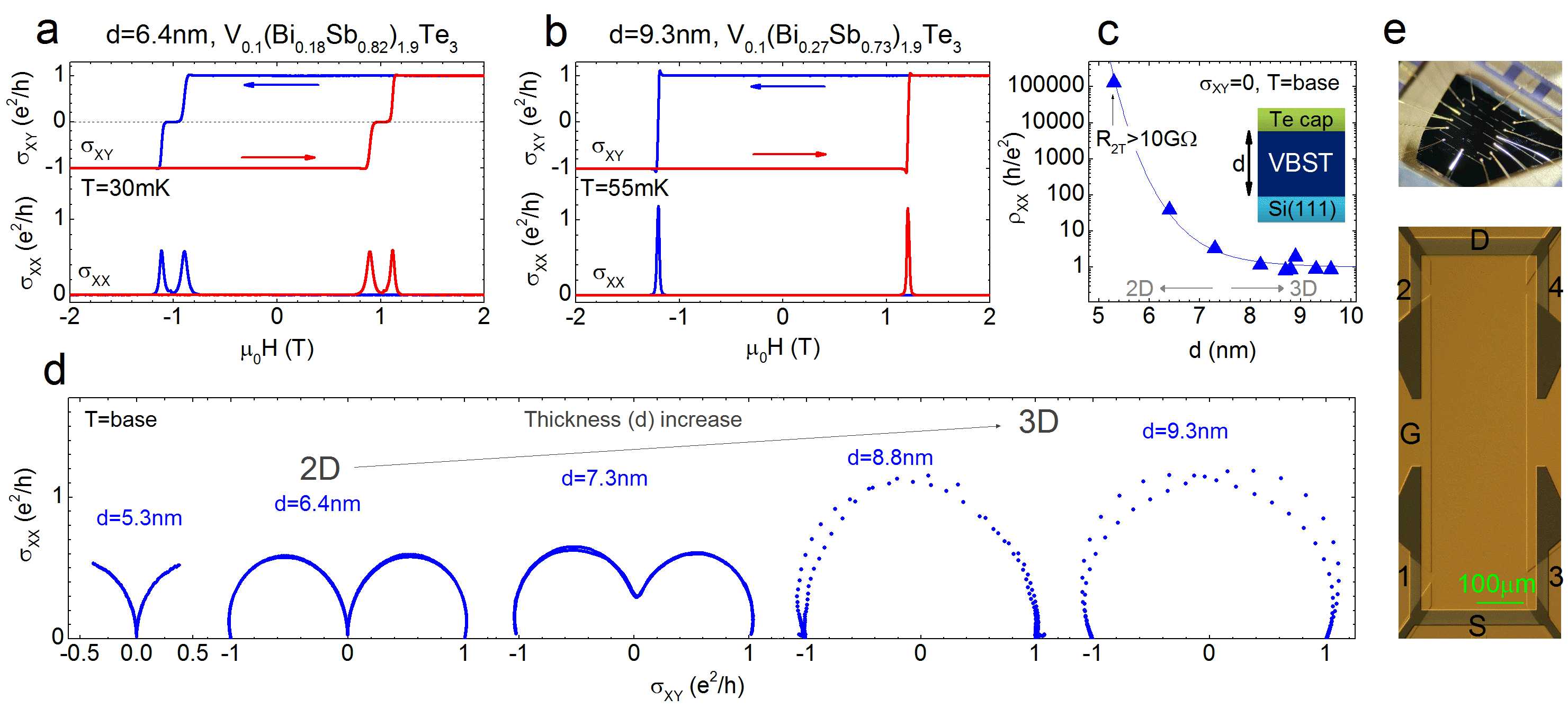}
\caption{
a) External magnetic field dependence of the conductivity tensor elements for a device patterned from a 6.4 nm thick $V_{0.1}$(Bi$_{0.18}$Sb$_{0.82}$)$_{1.9}$Te${_3}$ film at temperature 30 mK. b) same, for a 9.3nm thick $V_{0.1}$(Bi$_{0.27}$Sb$_{0.73}$)$_{1.9}$Te${_3}$ film at temperature 55 mK. c) Base temperature thickness dependence of the sheet resistance during the magnetization reversal at $\sigma_{xy}$=0 for all samples. VBST stands for (V,Bi,Sb)$_{2}$Te${_3}$, and the solid line is a guide to the eye. The sheet resistance for the 5.3 nm thick film is an estimated lower boundary (see Suppl. material~\cite{Suppl} for more details). d) Thickness evolution of the scaling behavior for samples ranging in thickness from the 2D integer quantum Hall to the 3D limit. e) Photograph of a patterned sample glued and bonded to a chip carrier (top), and optical microscope image of one of the devices (bottom) with labelled contacts.
}%
\label{fig:Fig1}%
\end{figure}
\noindent\rule[0.5ex]{\linewidth}{1pt}
\twocolumngrid

We approach the task of resolving the axion term from a different angle. We start from films of the TI (Bi,Sb)${_2}$Te${_3}$, where we can vary the thickness and thus ensure a gradual experimental transition between the 2D and 3D limits. When magnetic vanadium doping is used to gap out the surface states, this material system is established to show the quantum anomalous Hall effect.~\cite{Liu2008,Qiao2010,Yu2010,Nomura2011,Chang2013,Checkelsky2014,Kou2014,Chang2015,Bestwick2015,Grauer2015,Kou2015,Grauer2017} Our goal is to identify a platform where the combined Hall and longitudinal transport signals reveal the presence of distinct $e^2/2h$ surface contributions, and thus the quantized topological magneto-electric effect. We show that only a ($\sigma_{xy},\sigma_{xx}$) scaling diagram fingerprint can properly differentiate between a two-dimensional quantum anomalous Hall response~\cite{Wang2014,Checkelsky2014,Kou2015} and a truly three-dimensional axion Hall behaviour~\cite{Nomura2011,Grauer2017}. In particular, while maintaining perfectly quantized transport properties, we observe a scaling diagram transition from 2D quantum Hall type to 3D axion type of response as we increase the layer thickness. While the unprocessed magneto-transport data from all other axion insulator candidates in the literature show scaling indistinguishable from a 2D quantum anomalous Hall state, for sufficiently thick films our platform directly yields a 3D flow diagram.

Our (V,Bi,Sb)$_{2}$Te${_3}$ layers are grown using molecular beam epitaxy (MBE)~\cite{Winnerlein2017}, and patterned using standard optical lithography into six-terminal, 3 to 1 aspect ratio Hall bars, fitted with a top gate. The quality of our growth and processing approach is verified by the fact that such samples exhibit exact quantization of the anomalous Hall effect to metrological precision~\cite{Goetz2018}.
As depicted on the optical image of one of the devices in Fig.1e, the voltage in our experiments is applied between source (S) and drain (D), and longitudinal and Hall voltage drops are probed using contacts 1 through 4, allowing the determination of resistivity and conductivity tensor elements. Applying a gate voltage allows tuning of the Fermi level within the band structure. All data in this paper is collected with an applied gate voltage corresponding to the quantized anomalous Hall signal.

\onecolumngrid
\noindent\rule[0.5ex]{\linewidth}{1pt}

\begin{figure}[h]
\includegraphics[width=\columnwidth]{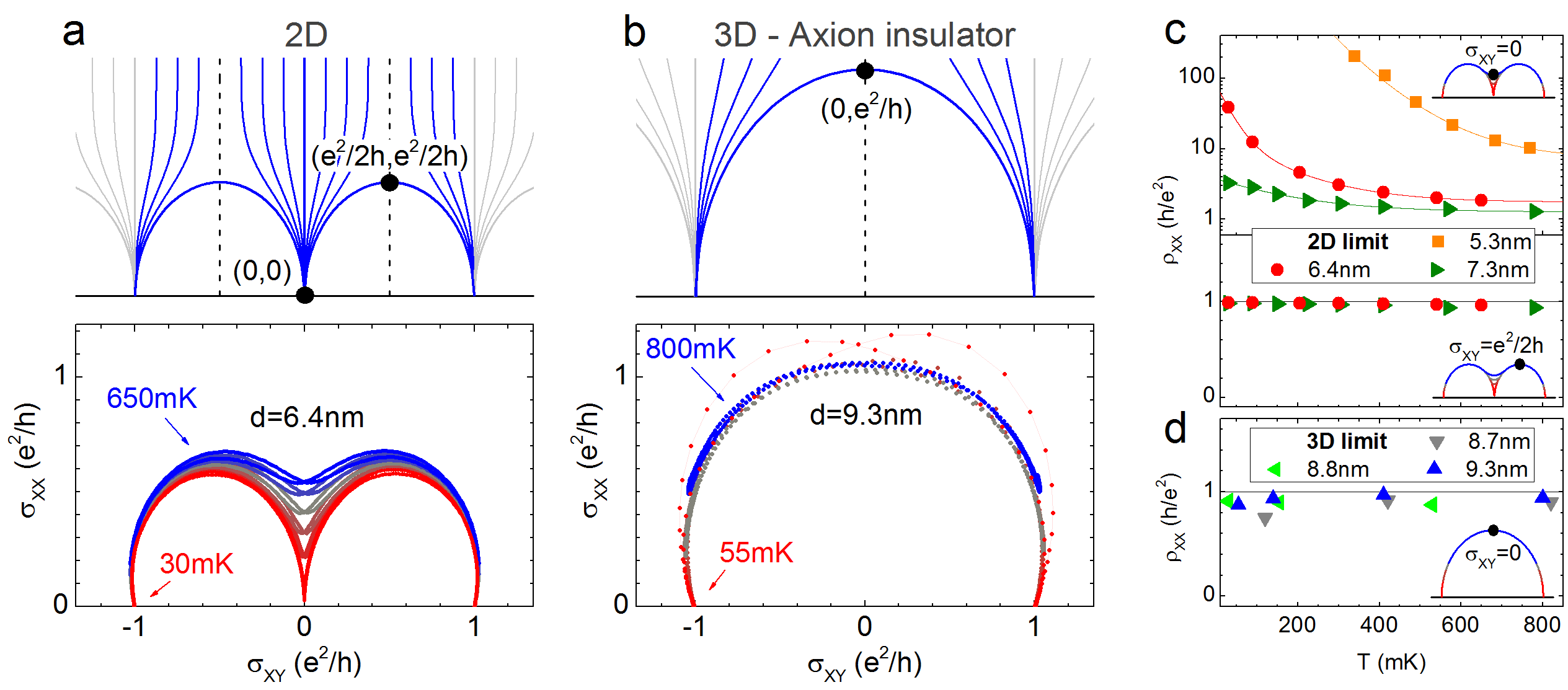}
\caption{
a) (top) Schematic of a global conductivity tensor flow diagram for the integer quantum Hall effect with dots marking the points used in the temperature dependence analysis. The scale on the axes matches the data panel below, (bottom) experimental magnetic field driven scaling diagram for a 6.4 nm thick film. The red color corresponds to a temperature of 30 mK and blue to 650 mK, with the intermediate values being 205 mK, 300 mK, 410 mK, and 540 mK. b) (top) Schematic of a global conductivity tensor flow diagram for Dirac fermions on two parallel topological interfaces, (bottom) experimental magnetic field driven scaling diagram for a 9.3 nm thick film. The red color corresponds to temperature 55 mK and blue to 800 mK, with intermediate values being 140 mK and 410 mK. c) Temperature dependence of film resistivity at $\sigma_{xy}$=0  and $\sigma_{xy}$=e$^2$/2h  for the layers in the 2D limit. The data for the 5.3 nm thick film is extracted from a two-terminal resistance measurement (see supplementary material for details). d) Temperature dependence of film resistivity when $\sigma_{xy}$=0 for the layers in the 3D axion limit. Solid lines in (c) and (d) are guides to the eye. The schematics inset to panel c) and d) indicate with a dot the point in the scaling diagram that is being plotted in the main figure.
}%
\label{fig:Fig2}%
\end{figure}
\noindent\rule[0.5ex]{\linewidth}{1pt}
\twocolumngrid

In Fig.1a we plot magneto-conductivity data collected from a sample patterned from a 6.4 nm thick V$_{0.1}$(Bi$_{0.18}$Sb$_{0.82}$)$_{1.9}$Te${_3}$ film measured at base temperature of a dilution refrigerator. The sample hosts a precisely quantized quantum anomalous Hall state. It is evident that when an external out of plane magnetic field is applied in order to reverse the magnetization in the film, a two-stage transition is observed with a well pronounced zero Hall plateau in $\sigma_{xy}$ and a double peak structure in $\sigma_{xx}$. Both conductivity components dropping to zero at the coercive field are a consequence of the sample going to high resistance during the plateau to plateau transition.  We also note in passing that such thin 2D films also have equivalent minor loop properties to magnetic trilayers (see suppl. material~\cite{Suppl} for minor loop measurement), such minor loops have been claimed to be a signature of independent switching of the two surface components, and thus of the antiparallel magnetization alignment in heterostructures~\cite{PhysRevLett.120.056801,tokura1,Mogi20172}.

Fundamentally different behavior is observed when the film thickness is increased. In Fig.1b we plot the corresponding magneto-conductivity data for a 9.3 nm thick V$_{0.1}$(Bi$_{0.27}$Sb$_{0.73}$)$_{1.9}$Te${_3}$ film also exhibiting a perfect quantum anomalous Hall state. During the out of plane magnetic field driven plateau to plateau transition a single well pronounced peak is observed in $\sigma_{xx}$ in absence of a zero Hall plateau in $\sigma_{xy}$, in sharp contrast to the thinner film. The finite value of $\sigma_{xx}$ when $\sigma_{xy}=0$ implies the absence of a high resistance phase, and in fact metallic behavior. Fig.1c gives the sheet resistance as a function of layer thickness during the plateau to plateau transition, when $\sigma_{xy}=0$ for a number of layers with thicknesses varying from 5.3 nm to 9.6 nm. The transition between an insulating and a metallic state is apparent, with sheet resistance changing by at least 5 orders of magnitude in this thickness range. The scaling behavior of $\sigma_{xx}$ vs $\sigma_{xy}$ during a transition between plateaus can be generated from the type of magneto-conductance data shown in Fig.1 a and b. Such scaling diagrams for assorted layers of different thicknesses are shown in Fig 1d. A smooth evolution of the scaling between two fundamentally distinct behaviors is observed.

It is well established that for the ordinary integer (and fractional) quantum Hall effect, this scaling follows a semicircular relation connecting the integer quantized values of $\sigma_{xy}$ (in case of the integer quantum Hall effect), with the center position of the semicircles occurring at ($\sigma_{xy}$,$\sigma_{xx}$)=((n+1/2)e$^2$/h,0) where n is an integer~\cite{Dykhne1994,Ruzin1995,Hilke1998,Hilke1999}. Phenomenologically, this has been conceptualized by modelling the plateau transition as a first-order transition between disordered quasiparticle excitations on top of the quantum Hall plateau at lower magnetic filling versus quasihole excitations on top of the one at higher filling~\cite{Dykhne1994}. From \cite{Dykhne1994,Ruzin1995,Hilke1998,Hilke1999}, we assume a scaling behavior equivalent to the integer quantum Hall effect to imply 2D physics, with ordinary Maxwell electrodynamics governing the system properties. This behavior was also previously reported in thin films of magnetic TIs exhibiting the quantum anomalous Hall effect~\cite{Wang2014,Checkelsky2014,Kou2015} and is clearly visible in our 6.4 nm thick film. It is worth noting that in the 2D limit, we find that the peak two-terminal resistance of the 5.3 nm thick layer exceeds 10G$\Omega$ (an estimated lower bound on the sample resistance, see suppl. material~\cite{Suppl} for more details), which is one order of magnitude larger than the value recently reported in a magnetic heterostructure (trilayer), and  attributed to unconventional insulating properties providing evidence of an axion insulator~\cite{Wu2020}.


\begin{figure}
\includegraphics[width=\columnwidth]{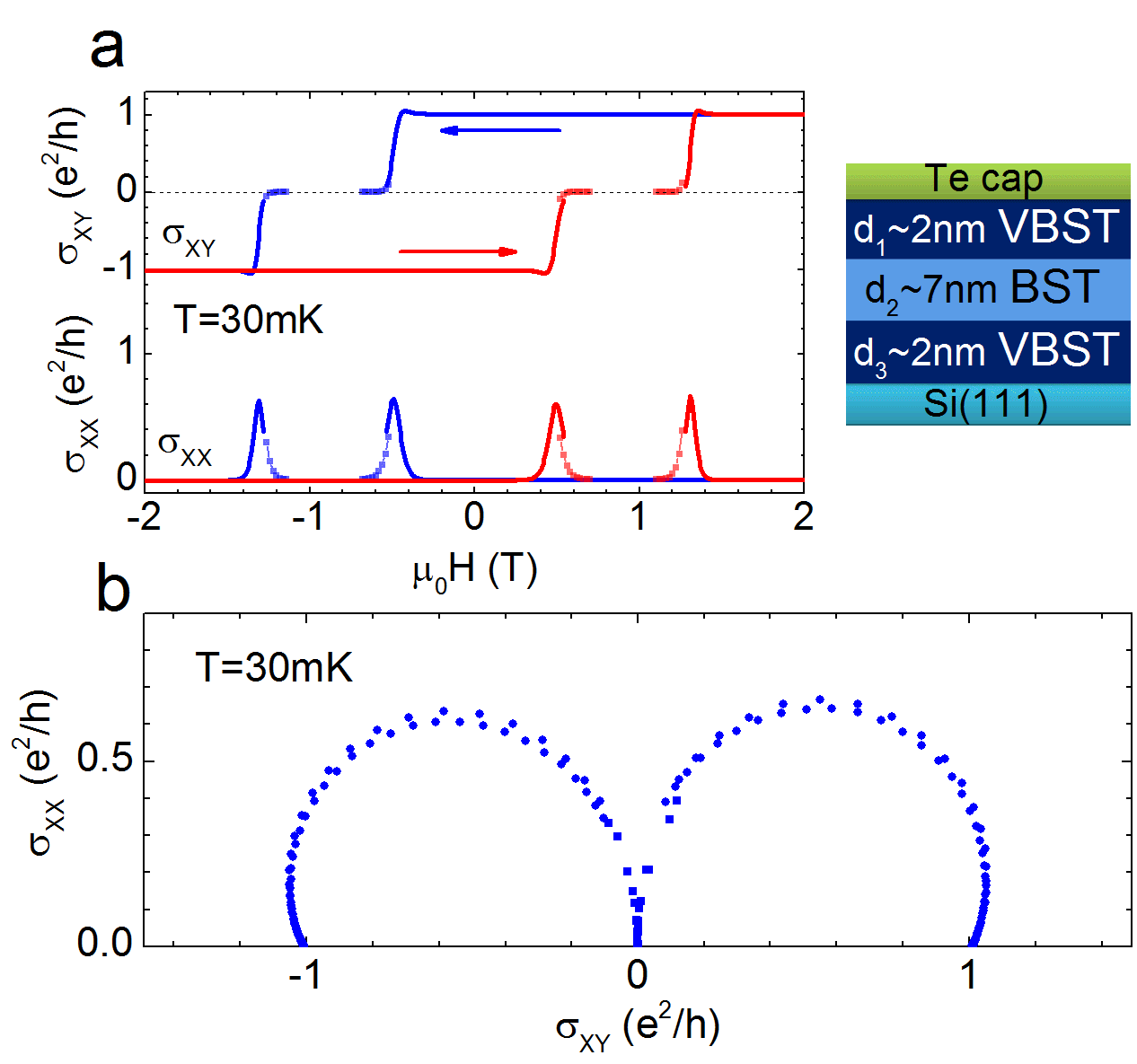}
\caption{
a) Hall and longitudinal magneto-conductivity data collected at a temperature of 30 mK for a sample patterned from a magnetic trilayer. A schematic of the hetero-structure is given on the right. VBST stands for V$_{0.1}$(Bi$_{0.31}$Sb$_{0.69}$)$_{1.9}$Te${_3}$, and BST stands for (Bi$_{0.31}$Sb$_{0.69}$)$_{2}$Te${_3}$. The plateau region where the sample resistance exceeds 10M$\Omega$ cannot be resolved in this measurement configuration (see suppl. material~\cite{Suppl} for the collected resistivity values). b) The conductivity tensor scaling behavior of the data in (a).
}%
\label{fig:Fig3}%
\end{figure}

As evidenced in Fig.1d, when the film thickness is increased a different scaling behavior is observed. The direct semicircular trajectory connecting the points ($\sigma_{xy}$,$\sigma_{xx}$)=($\pm$ e$^2$/h,0), is now centered at the origin (0,0). Nomura and Nagaosa investigated theoretically the scaling properties of a 2D topological state on the surface of a cylinder with magnetization normal to its surface, and enclosing a volume supporting axion electrodynamics~\cite{Nomura2011}. The scaling diagram they found consists of a semicircular trajectory directly connecting the points ($\sigma_{xy}$,$\sigma_{xx}$)=($\pm$ e$^2$/2h,0). This shift in position of the semicircle center to the origin of the ($\sigma_{xy}$,$\sigma_{xx}$) plane is a direct signature of a topological half-integer quantized $e^2/2h$ contribution (the sign of $\sigma_{xy}$ is determined by the direction of the magnetization). In an experiment in a planar device geometry, a 2D flat surface state can be treated mathematically as the infinite radius limit of the Nomura-Nagaosa cylinder. "Top" and "bottom" surfaces of our bulk films are individually such planes, with the magnetization normal pointing inwards at one interface and outwards at the other. 

This experimental geometry thus allows one to directly observe a non-trivial joint response from both parallel topological planes because charge carriers on the "top" and on the "bottom" interfaces experience a Lorentz force in the same direction and are thus deflected towards the same edge of the Hall bar. Having two surfaces results in a net doubling of the conductivity values predicted by Nomura and Nagaosa. Critically, the position of semicircle center located at the origin is robust against the number of parallel planes. Therefore the most plausible explanation for the phenomenology that we observe is that the single semicircular trajectory centered around point (0,0) visible in the thicker limit of our layers originates from the half-integer surface Chern-Simons term yielding a quantized topological magneto-electric effect, i. e. an axion insulator state.

Further distinctions between the two limits are found when the temperature evolution of the scaling behavior is analyzed. In Fig.2a and b we plot in the upper row the (schematic) flow diagram, and below, the scaling diagram of the magnetic field driven plateau transition for various temperatures for films in the 2D (6.4 nm) and 3D (9.3 nm) limits. The temperature evolution is best appreciated by looking at the value of $\rho_{xx}$ at the three specific points identified with dots in the scaling diagrams at the top of Fig.2 a and b (and equivalently in the schematic insets of Fig 2c and d). In the 2D limit, the point (0,0) implies an insulating state, and indeed as plotted in Fig.2c, the temperature dependence of the longitudinal resistivity when $\sigma_{xy}=0$ is consistent with a thermally activated high resistance behavior. For the same samples, a different temperature dependence is found at $\sigma_{xy}$=$\pm$ e$^2$/2h (i.e. at the top of the semicircle). The longitudinal resistivity there is nearly temperature independent with a value close to h/e$^2$. This implies the existence of a critically delocalized electron state. In the 3D limit, the longitudinal resistivity is found to be nearly temperature independent and always close to h/e$^2$ when $\sigma_{xy}$=0. This implies that the critical point (e$^2$/2h,e$^2$/2h) is shifted to (0,e$^2$/h) upon the dimensional transition, naturally implying a fractional $\sigma_{xy}$ shift of the semicircle center to the origin. This demonstrates that the two limits belong to fundamentally different quantum anomalous Hall classes. The sharpness of this transition as a function of thickness might seem surprising, however given that a transition to the 2D limit is a consequence of the overlap of evanescent topological surface state wave functions~\cite{Lu2010}, a sharp (exponential) thickness dependence is expected. The thickness at which we observe the transition lies in the vicinity of the reported 2D-3D transition observed with different experimental methods on similar layers. This spans the range from 2 quintuple layers (QL) in Bi$_{2}$Te${_3}$~\cite{Li2010}, 4 QL in Sb$_{2}$Te${_3}$~\cite{Jiang2012}, to 6 QL in Bi$_{2}$Se${_3}$~\cite{ZhangY2010}. The 2 and 4 QL scales reported in Bi$_{2}$Te${_3}$ and Sb$_{2}$Te${_3}$ are smaller than our transition thickness which appears to be closer to 5-6 QL. This difference may result from transport experiment (electrical resistance) having a different degree of sensitivity to the 2D-3D transition than the ARPES used in~\cite{Li2010} and STM used in~\cite{Jiang2012}. A proper understanding of these differences will require further research and is beyond the scope of this research article.

Now that we have established the scaling behavior in both the 2D (Maxwell) and 3D (axion) limit, we turn to the magneto-transport properties of our trilayer sample, which also exhibits perfect quantization of its quantum anomalous Hall effect. The heterostructure is comprised of two 2 nm thick layers of V$_{0.1}$(Bi$_{0.31}$Sb$_{0.69}$)$_{1.9}$Te${_3}$ separated by a 7 nm thick (Bi$_{0.31}$Sb$_{0.69}$)$_{2}$Te${_3}$ layer. In Fig.3 we plot the magneto-conductivity data and scaling behavior of this sample, obtained at base temperature of a dilution refrigerator. The behavior is consistent with that previously reported by other groups in similar structures~\cite{tokura1,Mogi20172,PhysRevLett.120.056801}, and shows a well pronounced zero Hall plateau in $\sigma_{xy}$, a double peak structure in $\sigma_{xx}$, and a scaling behavior comprised of two semicircles centered at ($\pm$ e$^2$/2h,0). 

The overall similarity of the data in Fig.3  to the magneto-transport in our uniformly doped 2D layers is striking. In~\cite{PhysRevLett.120.056801}, such data is nonetheless argued to constitute evidence of axion physics, but the claim is based on circular reasoning. Specifically, it is argued that the contributions of the "top" and "bottom" surfaces can be separated to extract a scaling curve that can only be inferred, but not measured, and that has the form of a semicircle centered on the origin. However, in order to perform the extraction, one must assume half-integer quantization of the single surface conduction, which is the very fact one is aiming to prove. 
Additionally in~\cite{PhysRevLett.120.056801}, magnetic field microscopy (MFM) measurements are presented as an attempt to justify this assumption. These are however not well spatially revolved, either vertically or in-plane, and simply show a uniform magnetization at the transition. While this could originate from antiparallel top and bottom surface magnetization, it could equally well come from the sample breaking up in domains smaller than the spatial resolution of the instrument, or even the long range ferromagnetic order completely collapsing.  Indeed, far from two sharp switching events, the reported MFM data clearly shows that the magnetization reversal process in that sample occurs through domain formation. Moreover the transport experiment suffers from the same limitations, as there is no way to demonstrate that the top and bottom magnetic layers can be manipulated independently of each other. When grown separately, Cr and V doped (Bi,Sb)$_{2}$Te${_3}$ films have significantly different coercive fields, but there is no verification that this is true when the two layers interact in the same heterostructure. Naturally this problem is avoided in the uniformly doped single film geometry as in our 3D limit, where magnetization reversal at both "top" and "bottom" surfaces happens simultaneously, and a fractional half-integer semicircle scaling shift becomes a direct experimental observable, rendering the "top"-"bottom" distinction obsolete.

Note also that in a true insulator the so-called "Zero Hall plateau", be it in resistance or conductance, is a fundamentally invalid measurement. One should always remember that a non-local 4 terminal transport resistance does not measure any true resistance (or conductance), but rather a voltage that is created between two points in response to a current flowing between two other points. The ratio is referred to as a resistance since it has units of Ohm, but it does not have meaning outside of a measurement context. In the case of a Hall measurement, one passes a current through the sample and measures a voltage orthogonal to the current. As long as the sample is conducting, this is a valid measurement, but once the sample is insulating (i. e. has higher impedance than the input of the voltmeter/cryostat) this stops being meaningful.

For an insulating sample, no current can flow through it. If the sample was being voltage biased, the flowing current goes to zero, whereas if it was purely current biased, the current must flow through alternate paths such as the cryostat and instrument wiring. In either case, the voltage leads no longer see any potential, and indeed, for an insulating sample, are simply floating. In the best case, a Hall geometry measurement on an insulator would give an undefined resistance of 0V/0A. In practice, the actual value is entirely determined by circuit consideration and instrumental offsets.

Due to the additional structural complexity, magnetic hetero-structures introduce another source of ambiguities pertaining to the relevant thickness scales. Magnetic hetero-structures are typically assumed to be 3D based on the total layer stack thickness, which indeed would fall into a 3D regime when directly compared to our thickness analysis. Yet given the striking similarities between the magnetic hetero-structures and thin (2D) uniform films in all experimental aspects, it appears clear that the trilayers lack 3D bulk properties. A different thickness scale must therefore be determining the dimensionality of the system. The obvious candidate is the thickness of individual layers in the heterostructure stack. Indeed some of the authors of~\cite{PhysRevLett.120.056801} recently reported on observation of an increased amount of chiral edge states in alternating magnetic/nonmagnetic TI hetero-structures, where one of the parameters affecting the total Chern number is the thickness of individual layers in the stack~\cite{Zhao2020}. These results directly imply that the properties of individual layers in the stack determine the topological properties of the entire system, and there is no reason to believe otherwise in the case of the trilayers of refs.~\cite{PhysRevLett.120.056801,tokura1,Mogi20172}. All of the individual layers in the hetero-structure stacks reported in~\cite{PhysRevLett.120.056801,tokura1,Mogi20172} fall into a 2D regime if compared directly to our 2D-3D transition scale.
Moreover recent measurements of the temperature dependence in the same trilayers reveals a critical exponent consistent with that of a (2D) integer quantum Hall effect~\cite{Wu2020}. All of this implies that quantum Hall systems exhibiting the zero Hall plateau state have fundamentally 2D electrodynamic properties, thus lack a 3D bulk \emph{and cannot host axion properties}. This also applies to the magnetically doped trilayer structures. The 3D magnetic TI geometry which we use does, however, allow us to access the axion insulator state predicted by Nomura and Nagaosa~\cite{Nomura2011}, and clearly differentiate it from the 2D limit where an axion electrodynamic term cannot exist.

Our findings imply that the observation of a zero Hall plateau cannot be used as a sufficient indicator of a sample having axion physics properties, and thus that the evidence as presented in refs.\cite{tokura1,Mogi20172,PhysRevLett.120.056801,Liu2020} is insufficient to unambiguously distinguish itself from 2D behavior with classical electrodynamics. On the contrary, our flow diagram analysis provides a clear distinction between the 2D and 3D limits, which is directly accessible as an experimental observable. 

\begin{acknowledgments}
We gratefully acknowledge the financial support of the Free State of Bavaria (the Institute for Topological Insulators), DFG (SFB 1170, 258499086), W\"urzburg-Dresden Cluster of Excellence on Complexity and Topology in Quantum Matter (EXC 2147, 39085490), and the European Commission under the H2020 FETPROACT Grant TOCHA (824140).
\end{acknowledgments}

\bibliography{KMFijalkowski_etal_ArXiv}

\setcounter{figure}{0}
\renewcommand{\thetable}{S\arabic{table}}  
\renewcommand{\thefigure}{S\arabic{figure}} 

\bigskip

\huge
\centerline{\textbf{Supplementary material}}
\normalsize

\section{1. Sample fabrication and experimental details.}
(V,Bi,Sb)$_2$Te$_3$ layers were grown on a Si(111) substrate, using molecular beam epitaxy (MBE). The growth temperature for the (V,Bi,Sb)$_2$Te$_3$ film was 190 $^{\circ}$C. The Bi/Sb ratio was determined by X-ray diffraction (XRD) measurements of the lateral and out of plane lattice constants. For all layers the Bi/Sb ratio is close to 1/4 and the vanadium doping level is approx. $2~\%$ of all atoms (the detailed composition of all of the films is presented in Table S1), for optimal quantum anomalous Hall conditions. Selective magnetic doping in the magnetic trilayer was achieved by timed opening and closing of the vanadium cell shutter during the co-deposition of Bi, Sb, and Te. Layers were capped in-situ with a 8-10 nm thick insulating Te layer as a protection from ambient conditions and the lithographic process. The concentration of magnetic doping was determined from reference samples using energy-dispersive X-ray spectroscopy (EDX). The topological insulator film thickness was determined using X-ray reflectivity (XRR) measurements of the layers investigated in transport, as well as reference samples grown without a protective Te cap. The systematic uncertainty on the absolute value of the film thickness is about $\pm$1 nm. This is the result of some film roughness and XRR limitations in determining the position of (V,Bi,Sb)${_2}$Te${_3}$/Te interface. However, the relative thickness difference uncertainty between different investigated films is smaller, and estimated to be $\pm$0.3 nm.
\bigskip

After growth, the samples were patterned using standard optical lithography methods. The devices consist of an Ar plasma ion beam etched (IBE) mesa, covered with 20 nm of AlOx, 1 nm of HfOx, 2 nm of Ti and 100 nm of Au as a gate dielectric and electrode. Ohmic contacts are obtained by removal of the Te cap from the contact area with Ar plasma IBE, followed by deposition of a layer stack of 50 nm of AuGe and 50nm Au, without breaking the high vacuum conditions. Each sample contains two six-terminal Hall bar devices with widths 10 $\mu$m and 200 $\mu$m, respectively, and an aspect ratio 3/1. No qualitative difference is found between the sizes of Hall bars. Finished samples are glued to a chip carrier, and electrically connected using mechanically wedge bonded Au wires. In Table S1 we list the relevant details of all investigated devices.
\bigskip

Measurements are done using standard low excitation voltage (10-100$\mu$V) and low frequency AC (less than 20Hz) lock-in techniques, with the exception of the lower bound estimate for the resistance of 5.3 nm thick film at 60 mK,  where a square wave 2 mV, 5 mHz quasi-DC method is used. Measurements are performed in dilution refrigerator systems with external magnetic field applied perpendicular to the plane of the sample for Hall measurements. 
\bigskip

\begin{table}
\begin{tabular}{| c | c | c | c |}
\hline
thickness~"\textbf{d}" & Sb~"\textbf{x}" & Hall bar dimensions\\ \hline
\multicolumn{3}{| c |}{Uniformly doped layers:}\\ \hline
\textbf{5.3~nm} 	& 0.71 & 200x600 $\mu$m$^2$\\ \hline
\textbf{6.4~nm} 	& 0.82 & 200x600 $\mu$m$^2$\\ \hline
\textbf{7.3~nm} 	& 0.80 &	200x600 $\mu$m$^2$\\ \hline
\textbf{8.2~nm} 	& 0.80 & 10x30 	$\mu$m$^2$\\ \hline
\textbf{8.7~nm} 	& 0.73	& 200x600 $\mu$m$^2$\\ \hline
\textbf{8.8~nm} 	& 0.79 & 10x30 	$\mu$m$^2$\\ \hline
\textbf{8.9~nm} 	& 0.70 & 200x600 $\mu$m$^2$\\ \hline
\textbf{9.3~nm} 	& 0.73	& 200x600 $\mu$m$^2$\\ \hline
\textbf{9.6~nm} 	& 0.70 & 200x600 $\mu$m$^2$\\ \hline
\multicolumn{3}{| c |}{Magnetic trilayer:}\\ \hline
\textbf{2~nm/7~nm/2~nm} 	& 0.69 & 200x600 $\mu$m$^2$\\ \hline
\end{tabular}
\caption{Table presenting thicknesses and composition of individual V$_{0.1}$(Bi$_{1-x}$Sb$_x$)$_{1.9}$Te$_3$ layers of thickness "d", used to pattern the investigated devices, as well as the Hall bar dimensions for each device. The last row represents a V$_{0.1}$(Bi$_{1-x}$Sb$_x$)$_{1.9}$Te$_3$ / (Bi$_{1-x}$Sb$_x$)$_{2}$Te$_3$ / V$_{0.1}$(Bi$_{1-x}$Sb$_x$)$_{1.9}$Te$_3$ trilayer.
}%
\label{tab:TabS1}%
\end{table}

\section{2. Magneto-resistance measurements from the layers of different thicknesses at base temperature.}
In Fig. S1 we present the magneto-resistivity measurements from the samples of various thicknesses from the Fig.1 in the main text. It includes 9 separately grown films with thicknesses spanning the range from 5.3 to 9.6 nm. The measurements were collected at the lowest experimentally accessible temperature depending on the available dilution refrigerator. Values of the temperature and layer thickness are labeled in the Figure for each curve. The device patterned from the 5.3 nm thick film, exhibit a too high resistance during the magnetization reversal to reliably measure the resistivity tensor elements using a four-terminal configuration, data in that field range is thus inaccessible and was omitted in the data to obtain the scaling plot in Fig.~1d of the main text.
\bigskip

\section{3. Lower bound sheet resistance estimate for the thinnest layer at base temperature.}
In Fig. S2 we present the measurement used to determine a lower bound on the 5.3 nm thick layer sheet resistance. In order to eliminate the capacitive coupling between the leads in the cryostat, as well as the capacitive coupling between the gate electrode and the sample, we utilized a reversed DC (or quasi-DC) method. This approach allows one to significantly reduce the capacitive shunt effects, while also limiting the effect of experimental offsets such as thermal voltages in the cryostat and instrumental offsets. The measurement is performed with the sample at a temperature 60 mK, and the voltage excitation is a square wave with a frequency of approx. 5 mHz and an amplitude 2 mV. The circuit consists of a series 10 M$\Omega$ reference resistor in order to determine the total current value, placed on the voltage source side of the circuit (before the cryostat) (see Fig. S2e for a circuit diagram schematic). Due to possible alternative current paths to the ground, resulting from shunt effects in the cryostat, the equivalent resistance of the circuit (after the reference resistor) is a lower bound on the sample resistance, therefore we further attempt to estimate the equivalent circuit resistance. In Fig. S2a we plot the voltage drop over the reference resistor, as well as the equivalent voltage. The magnetic field is ramped at a constant rate of 5 mT per minute and the data is collected as a function of time. Our resistance estimate is a result of averaging over the range where the sample goes insulating (dashed lines in Fig. S2a). 
In Fig. S2b we plot a zoomed in single period measurement of the reference voltage, with the two colors representing the two directions of bias voltage. In order to avoid (as much as possible) capacitive effects resulting from the high frequency voltage component generated by the switching of the bias voltage direction, the first half of the data points after each reversal is removed from the analysis. To obtain a measure of the reference voltage amplitude, the difference between the consecutive pairs of points in each bias direction is calculated and divided by a factor of 2 (1st red point with 1st blue point, 2nd red point with 2nd blue point etc.) resulting in 25 data points from each excitation period for the statistical analysis. We plot a histogram representing the distribution of reference voltage values (Fig. S2c) and the corresponding equivalent voltage values (Fig. S2d) from 550 data points in the insulating plateau range. It is clear that the reference signal is noise limited, therefore as a conservative upper bound on the reference voltage signal value, we take a standard deviation of the distribution (the real signal is clearly smaller than the standard deviation, but how much smaller is hard to quantify). Considering the mean value of the equivalent voltage and standard deviation of the reference voltage, we obtain a lower bound on the equivalent circuit resistance (and therefore a lower bound on the sample resistance) of 10.77 G$\Omega$ (if the mean value of the signal on the reference resistor is taken to be exact, the value becomes 34.5 G$\Omega$). Our Hall bar has 3 squares, so the sheet resistivity $\rho_{xx}$ (using the 10.77 G$\Omega$ value) is approx. 3.6 G$\Omega$, or equivalently 140000~h/e$^2$. 

We refer to this value as a lower bound since the total approx. 10 G$\Omega$ resistance of the sample is not negligible compared to the resistance to ground of the cryostat leads (of order of 100 G$\Omega$). For this reason, what is measured is an equivalent resistance as conceptually depicted in the schematic of Fig. S2e, where additional parallel paths to ground through the cryostat may exist. Without knowing the exact distribution of these parasitic paths to ground, a more precise detailed circuit analysis is not possible, but what is clear is that the equivalent resistance that we measure must be smaller (or equal to) the true resistance of the sample.
\bigskip

\section{4. Magneto-resistance data from the temperature series for samples of different thicknesses.}
In Fig. S3 we plot experimental magneto-resistivity data from which the data plotted in the Fig. 2 in the main text was obtained. The dataset contains magnetic field sweeps at different temperature values from three samples in the 2D limit, and three samples in the 3D limit.
\bigskip

\section{5. Magneto-resistance measurements from the magnetic trilayer at base temperature.}
In Fig. S4 we present the experimental magneto-resistivity values collected from the magnetic trilayer at base temperature, from which the data plotted in Fig. 3 of the main text was obtained. Total voltage excitation is the same for both measurements (100 $\mu$V). The measurement in Fig. S4(a) is performed at 13.7 Hz voltage excitation frequency with a series 9.9 k$\Omega$ reference resistor, in order to capture the quantum anomalous Hall plateau regime. The measurement in Fig. S4(b) is performed at 2 Hz frequency with a series 997 k$\Omega$ reference resistor, in order to better resolve the insulating regime during the plateau to plateau transition. The unusual shape of $\rho_{xx}$ and $\rho_{yx}$ during magnetization reversal is the consequence of capacitive coupling between the leads in the cryostat affecting the current flow in the circuit when the sample goes insulating, as well as the reference voltage dropping nearly to zero resulting in a division by small number when calculating the resistances. In the highly insulating regime the four-terminal measurements are no longer valid, therefore the relevant range was omitted from the data plotted in the main text.
\bigskip

\section{6. Minor loop for a 2D layer at base temperature.}
In Fig. S5 we plot magneto-conductivity, magneto-resistivity, and raw voltages from the data obtained from a uniformly doped 2D 6.4 nm thick layer. The red color represents a regular full magnetic field range sweep in one direction, whereas the blue color represents a minor loop measurement. A clear minor loop similar to those observed in trilayers and presented as proof of independent switching of the two layers is observed in this homogenous 2D layer.  In Fig. S5(c) we plot a corresponding scaling behavior obtained from the data in (a) and (b), confirming that no hint for axion physics emerges.

\onecolumngrid

\begin{figure}
\includegraphics[width=\columnwidth]{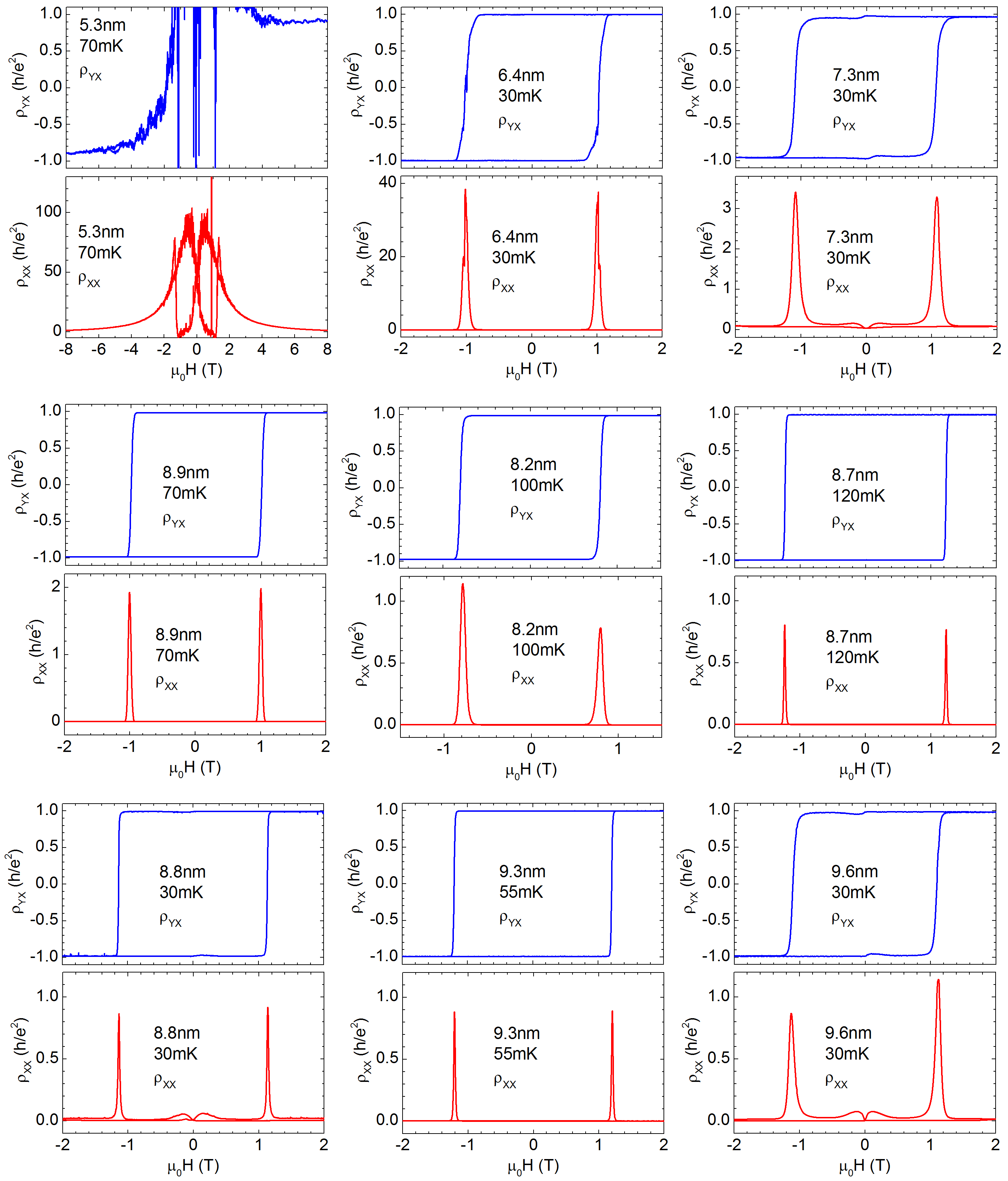}
\caption{
Magneto-resistivity measurements collected from 9 samples of various thicknesses, at experimental base temperature. The measurements correspond to the analysis in Fig.1 of the main text. Every curve is labeled accordingly in the figure. The relevant parameters for each sample are summarized in Table S1.
}
\label{fig:Fig1}%
\end{figure}
\twocolumngrid

\onecolumngrid

\begin{figure}
\includegraphics[width=\columnwidth]{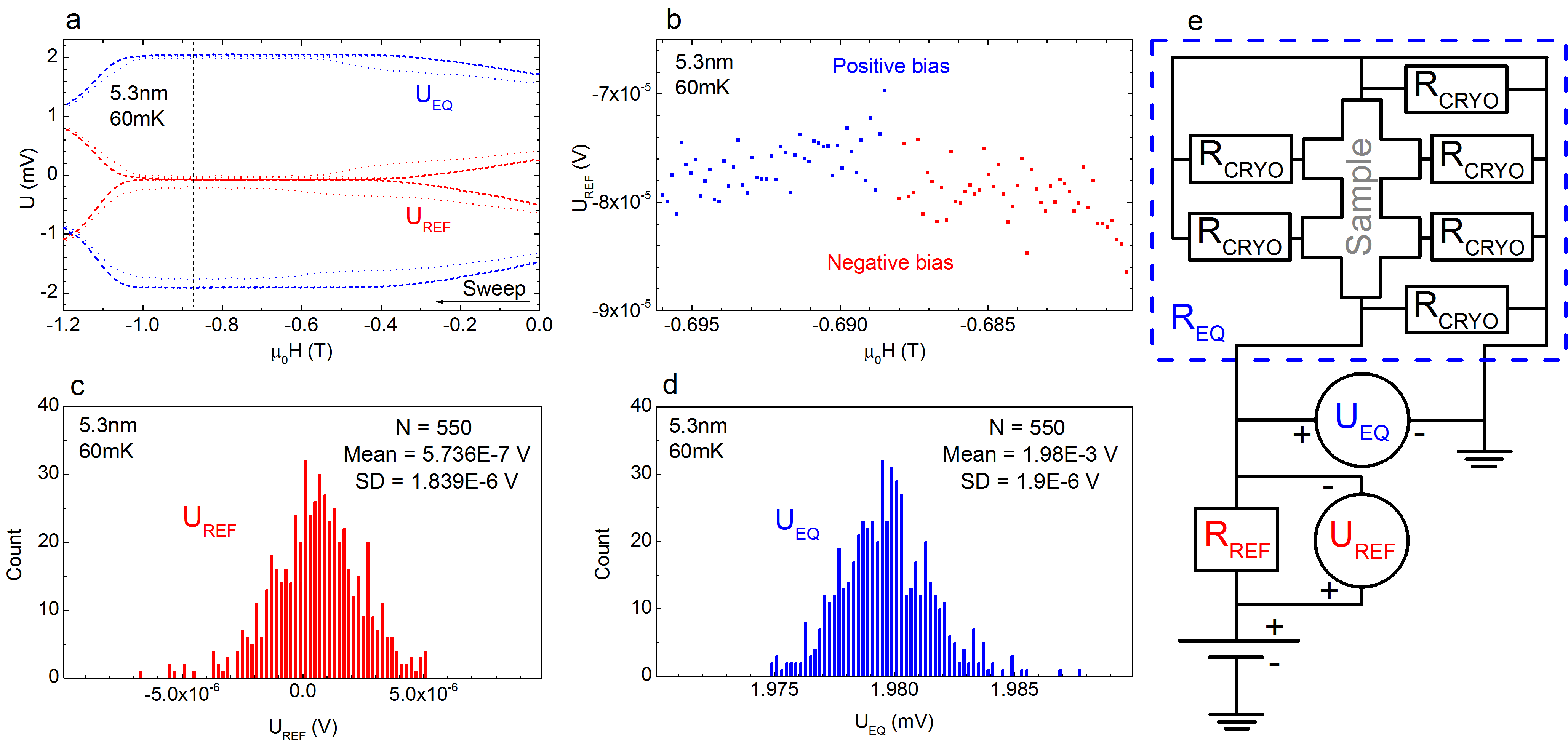}
\caption{
Estimate of the lower bound of resistance of a sample patterned from a 5.3nm thick film. The measurement was performed at temperature 60 mK, with a square wave excitation of 5 mHz frequency and 2 mV amplitude. a) Equivalent voltage of the circuit (blue points), and voltage drop over the series 10 M$\Omega$ reference resistor (red points). Dashed lines represent the insulating range used for counting statistics. b) Zoomed in section of the reference voltage covering a single period in the insulating regime, blue and red points correspond to the two bias directions. c) and d) Histogram counting of the obtained voltage drop over the reference resistor (c) and equivalent voltage (d). Counting statistics covered 550 data points, and obtained mean values and standard deviations are labeled accordingly in the figures. (e) Schematic of a circuit diagram. R$_{REF}$ is a reference 10 M$\Omega$ resistor, R$_{CRYO}$ represents the unintentional shunt effects in the cryostat wiring, R$_{EQ}$ shows equivalent resistance of the circuit (dashed box), U$_{REF}$ and U$_{EQ}$ represent voltage drops over the reference resistor and equivalent resistor.
}
\label{fig:Fig1}%
\end{figure}
\twocolumngrid

\onecolumngrid

\begin{figure}
\includegraphics[width=\columnwidth]{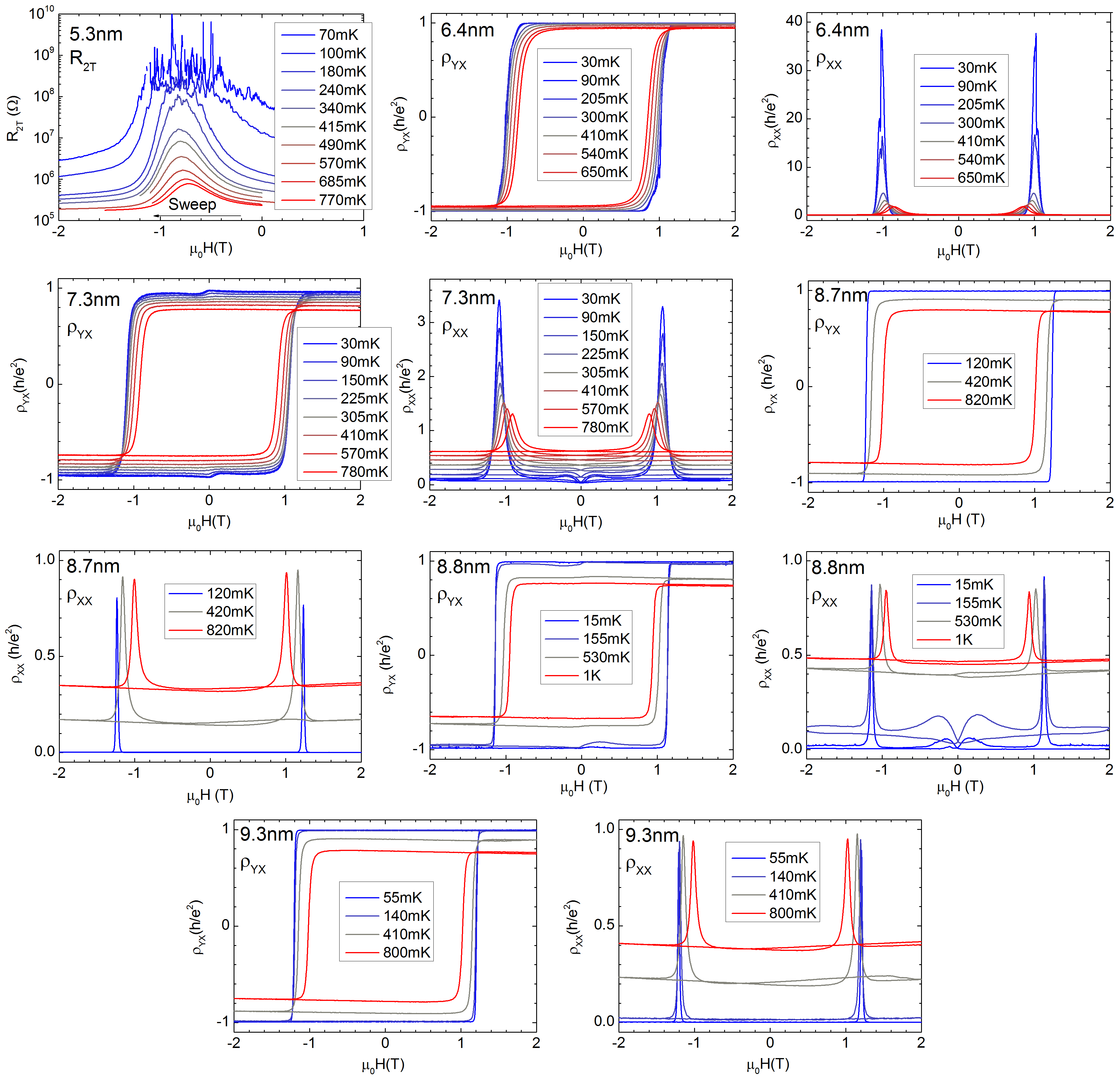}
\caption{
Magneto-resistivity data for a wide range of fixed temperature values, for all samples analyzed in Fig.2 of the main text. For the most insulating 5.3 nm thick film we analyze a two-terminal resistance. 
}
\label{fig:Fig1}%
\end{figure}
\twocolumngrid

\onecolumngrid

\begin{figure}
\includegraphics[width=\columnwidth]{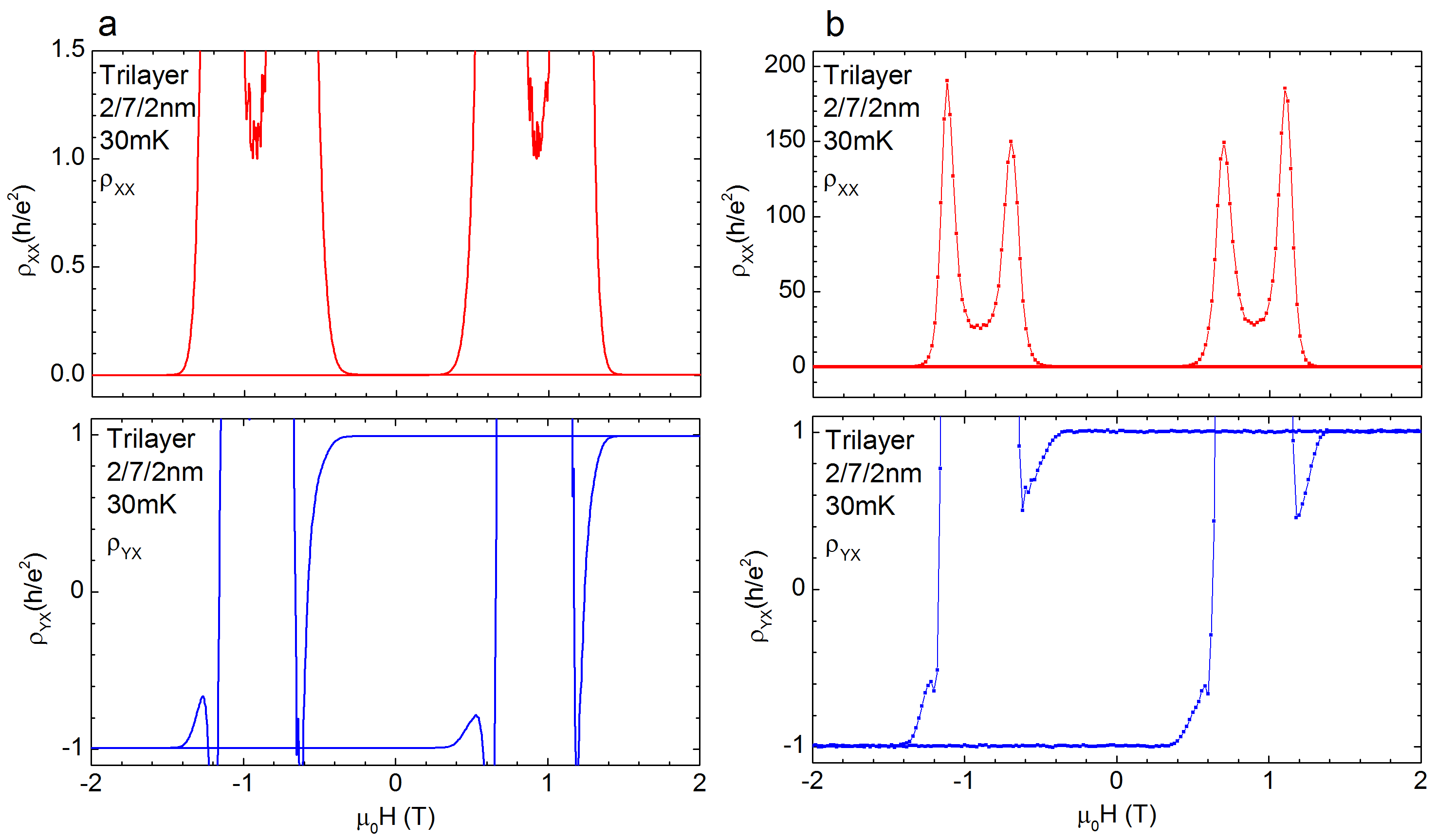}
\caption{
Magneto-resistivity values for a magnetic trilayer at base temperature. Data corresponds to the data plotted in the Fig. 3 of the main text. Total excitation voltage is 100 $\mu$V for each measurement. The measurement in (a) is at 13.7 Hz excitation voltage frequency with a series 9.9 k$\Omega$ reference resistor, in order to capture the quantum anomalous Hall plateau regime. Measurement in (b) is at a frequency of 2 Hz with a series 997 k$\Omega$ reference resistor, in order to better resolve the high resistance regime during the plateau to plateau transition.
}
\label{fig:Fig1}%
\end{figure}
\twocolumngrid

\onecolumngrid

\begin{figure}
\includegraphics[width=\columnwidth]{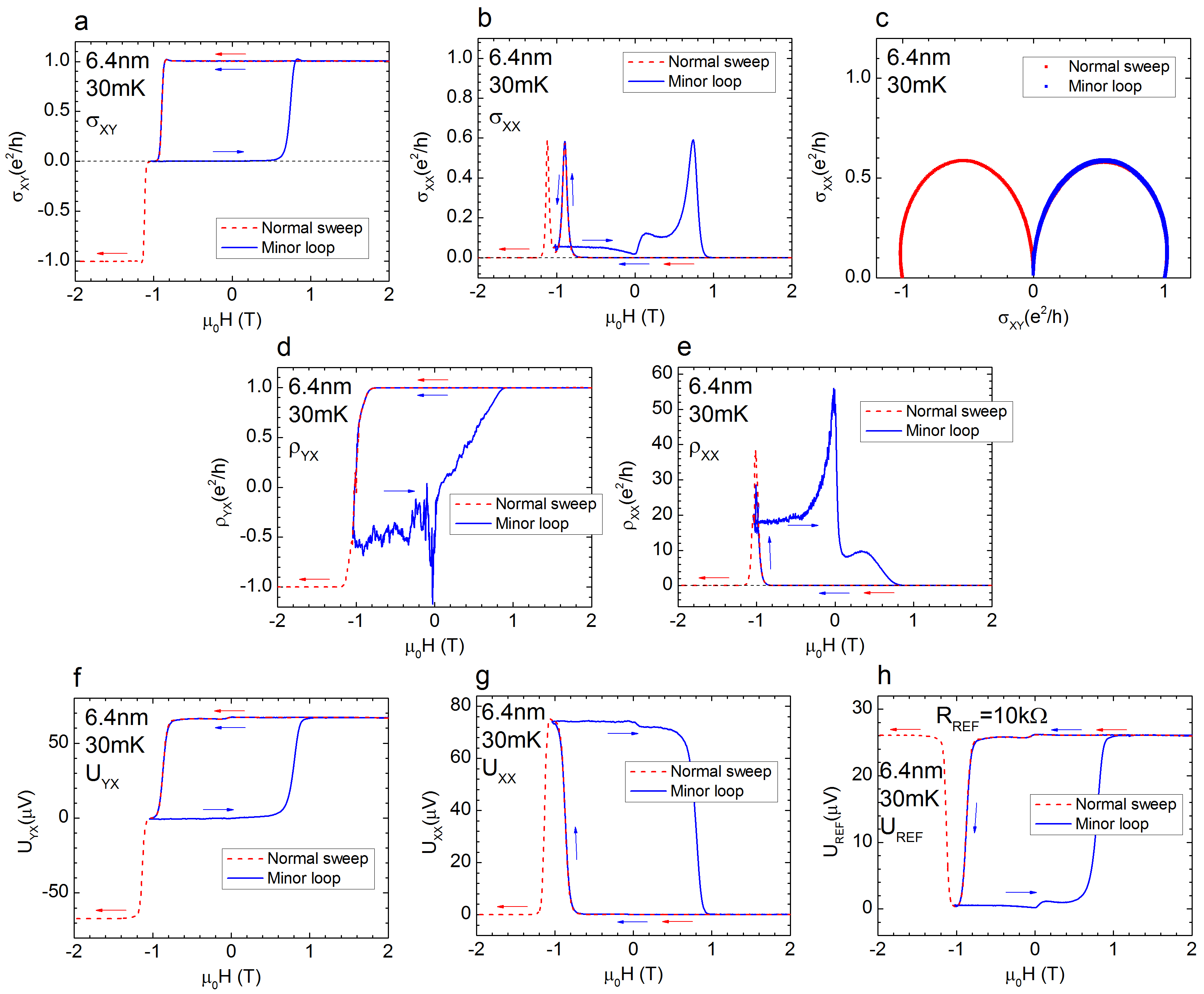}
\caption{
Minor loop transport data for a 6.4 nm thick film. a) Hall conductivity data representing a full magnetic field sweep in one direction (red color), and a minor loop (blue color). b) Longitudinal conductivity representing a full magnetic field sweep in one direction (red color), and a minor loop (blue color). c) Scaling behavior calculated from the data in (a) and (b). d) and e) Corresponding Hall and longitudinal resistivity data. f-h) Corresponding raw voltage data from which the resistivity and conductivity tensor elements were calculated. U$_{REF}$ represents a voltage drop over a series reference resistor, used to determine the current. Colored arrows show the sweep direction.
}
\label{fig:Fig1}%
\end{figure}
\twocolumngrid

\end{document}